\documentclass{PoS}
\usepackage{epsfig,cite}

\title{
\vspace*{-2.9cm}
\begin{minipage}{\textwidth}
{\normalfont\small LTH 1010, DESY 14-068, Nikhef 2014-011, LPN 14-071
\hspace{\fill} May 2014}\\
\end{minipage}\\[60pt]
  A calculation of the three-loop helicity-dependent splitting functions in QCD}

\ShortTitle{The three-loop helicity-dependent splitting functions}

\author{\speaker{A. Vogt}\\
        \mbox{Department of Mathematical Sciences, University of Liverpool,
        Liverpool L69 3BX, UK}\\
        E-mail: \email{Andreas.Vogt@liv.ac.uk}}

\author{S. Moch \phantom{g}\\
        \mbox{II.~Institut f\"ur Theoretische Physik, Universit\"at Hamburg,
        D-22761 Hamburg, Germany}\\
        E-mail: \email{sven-olaf.moch@desy.de}}

\author{J.A.M. Vermaseren \phantom{g}\\ 
        \mbox{NIKHEF Theory Group, Science Park 105, 1098 XG Amsterdam,
        The Netherlands}\\
        E-mail: \email{t68@nikhef.nl} \\ \\ }

\abstract{
We have calculated the complete matrix of three-loop helicity-difference
(`polarized') splitting functions $\Delta P^{\,(2)}_{ik}(x)$, $i,k= \rm q,g$, 
in massless perturbative QCD. In this note we briefly discuss some properties 
of the polarized splitting functions and our non-standard determination of the 
hitherto missing lower-row quantities $\Delta P^{\,(2)}_{\rm gq}$ and 
$\Delta P^{\,(2)}_{\rm gg}$. The resulting next-to-next-to-leading order 
(NNLO) corrections to the evolution of polarized parton distributions are 
illustrated and found to be small even at rather large values of the strong 
coupling constant $\as$.
}

\FullConference{Loops and Legs in Quantum Field Theory\\
                27 April 2014 - 02 May 2014, Weimar, Germany}

\usepackage{color,colordvi}
\def\colour4colour#1{\Blue{#1}}

\def\ca{{C_{\!A}}}
\def\cas{{C^{\, 2}_{\!A}}}
\def\cat{{C^{\, 3}_{\!A}}}
\def\cf{{C_F}}
\def\nf{{n^{}_{\! f}}}
\def\nfs{{n^{2}_{\! f}}}

\def\cfs{{C^{\, 2}_{\! F}}}
\def\cft{{C^{\, 3}_{\! F}}}

\def\DNnO{D_0}
\def\DNmO{D_{-1}}
\def\DNpO{D_{1}}
\def\DNppO{D_{2}}
\def\DNn#1{D_0^{\,#1}}

\def\DNp#1{D_1^{\,#1}}

\def\z#1{{\zeta_{#1}^{}}}
\def\zs2{{\zeta_{2}^{\,2}}}
\def\zt2{{\zeta_{2}^{\,3}}}
\def\zf2{{\zeta_{2}^{\,4}}}

\def\S(#1){{{S}_{#1}}}
\def\Ss(#1,#2){{{S}_{#1,#2}}}
\def\Sss(#1,#2,#3){{{S}_{#1,#2,#3}}}
\def\Ssss(#1,#2,#3,#4){{{S}_{#1,#2,#3,#4}}}

\def\frct#1#2{\mbox{\Large{$\frac{#1}{#2}\,$}}}

\newcommand{\beq}{\begin{equation}}
\newcommand{\eeq}{\end{equation}}
\newcommand{\bea}{\begin{eqnarray}}
\newcommand{\eea}{\end{eqnarray}}
\newcommand{\nn}{\nonumber}

\newcommand{\ra}{\rightarrow}
\newcommand{\ars}{a_{\sf s}}
\newcommand{\ass}{\alpha_{\sf s}}
\newcommand{\as}{\alpha_{\sf s}^{}}
\newcommand{\ar}{a_{\sf s}^{}}
\newcommand{\ep}{\varepsilon}

\newcommand{\MSbar}{$\overline{\mbox{MS}}$}

\begin{document}


\section{Introduction: Polarized PDFs, their evolution, 
{\normalsize $\ass^2$} calculations, large-x limit}

\noindent
\vspace*{-1mm}
The unpolarized and polarized parton distributions of a longitudinally 
polarized hadron are given~by$\!\!$ 
\beq
  f_{i}^{}(x,\mu^2) \:=\:
  f_{i}^{\,\ra}(x,\mu^2) \,+\, f_{i}^{\,\leftarrow}(x,\mu^2)
  \quad \mbox{ and } \quad
  \Delta f_{i}(x,\mu^2) \:=\:
  f_{i}^{\,\ra}(x,\mu^2) \,-\, f_{i}^{\,\leftarrow}(x,\mu^2) 
  \; ,
\eeq
respectively, in terms of the quark and gluon distributions $\,f_{i}^{\,\ra}$ 
and  $f_{i}^{\,\leftarrow}$ for the same and opposite helicity. Here $x\,$
is the parton's momentum fraction, and $\mu\,$ denotes the 
factorization scale which, in the present context, can be identified with the 
renormalization scale without loss of information.

\vspace*{0.5mm}
Their scale dependence is governed by the renormalization-group evolution
equations
\beq
  \frac{d}{d \ln \mu^2} \, (\Delta) f_i^{}(x,\mu^2) 
  \; =\; \left[ {(\Delta) P^{}_{ik}(\as(\mu^2))} 
  \:\!\otimes\:\! (\Delta) f_k^{}(\mu^2) \right]\! (x) \; ,
\eeq
where $\otimes$ represents the standard Mellin convolution. The expansion
of the respective splitting functions powers of the strong coupling constant 
$\as(\mu^2)$ can be written as
\beq
\label{Pexp}
  (\Delta) P^{}_{ik}(x,\mu^2) \;=\;
  {\textstyle \sum_{\,n=0}} \; \ars^{\,n+1} (\Delta) P^{\,(n)}_{ik}(x)
  \quad \mbox{ with } \quad
  \ars^{} \:\equiv\: \as(\mu^2)/(4\:\!\pi) \; .
\eeq
The third-order (NNLO) contributions $\Delta P^{\,(2)}_{ik}$ for the polarized 
case are the subject of this note.


\vspace*{2mm}
The corresponding second-order order calculations were performed in the 1990s,
when a lot of attention was devoted to the polarized parton distributions in 
the wake of the `spin-crisis' set off by Ref.~\cite{EMCpol1} in 1988. 
All these calculations were performed in the framework of dimensional 
regularization, and thus had to address the treatment of the Dirac matrix 
$\gamma_{5}^{}$ in $D \neq 4$ dimensions.

\vspace*{0.5mm}
The splitting functions $\Delta P^{\,(1)}_{\rm qq}$ and 
$\Delta P^{\,(1)}_{\rm qg}$ were obtained, together with the second-order 
coefficient functions for the structure function $g_1^{}$ in polarized 
deep-inelastic scattering (DIS) by Zijlstra and van Neerven in 1993 
\cite{ZvNpol}, using the so-called Larin scheme \cite{g5-Larin} with
$
  p\!\!\!\!\!\,/ \:\! \gamma_{5,L}^{} \:=\: \frac{i}{6}\:
  \ep_{p\mu\nu\rho}\, \gamma_\mu \gamma_\nu \gamma_\rho \, ,
$
where the resulting contractions of the $\ep$-tensor are evaluated in terms 
of the $D$-dimensional metric.

\vspace*{0.5mm}
The complete matrix $\Delta P^{\,(1)}_{ij}$ was calculated in 1995 
independently by Mertig and van Neerven \cite{MvN} and by Vogelsang 
\cite{WVdP1}. The former calculation was performed in the framework of the
operator product expansion (OPE) and used the `reading-point' scheme for 
$\gamma_{5}^{}$ \cite{g5-readp}. The latter calculation was carried out
in the lightlike axial-gauge approach and employed primarily the 
`t Hooft/Veltman prescription for $\gamma_{5}^{}$ of Refs.~\cite{g5-HVBM}
which, in the present context, is equivalent to the Larin scheme.

\vspace*{0.5mm}
The relation of the prescriptions of Refs.~\cite{g5-Larin,g5-HVBM} to the 
\MSbar\ scheme was addressed to second order (NNLO) in 1998 in 
Ref.~\cite{MSvN}, where the transformation matrix is of the form
\beq
  Z_{ik}(\as(\mu^2)) \:=\: \delta_{\,iq} \delta_{\,kq} \big(
     \ar \,  z_{\rm ns}^{\,(1)} 
   + \ars^2  \big[ z_{\rm ns}^{\,(2)} 
     + z_{\rm ps}^{\,(2)} \big]
   + \,\ldots \big) \; .
\eeq
Its non-singlet (ns) entries can be fixed by the relation between the 
corresponding coefficient functions for $g_1^{}$ and the structure function
$F_3^{}$ which is known to order $\ass^{\,3}$ \cite{MVV10}. The critical 
part is thus the pure-singlet (ps) part for which only that one calculation has
been performed so far.  


\vspace*{2mm}
For reasons that will become obvious below, it is important for us to control
the $x\!\ra\! 1$ threshold limits of the splitting functions. Here it is 
reasonable to expect a helicity-flip suppression by a factor of 
$(1\!-\!x)^2$ or $1/N^{\,2}$ in Mellin space, cf.~Ref.~\cite{BBSsuppr}.
E.g., the differences 
$\,
  \delta_{ik}^{\,(0)} \,\equiv\, P^{\,(0)}_{ik} - \Delta P^{\,(0)}_{ik}
$
of the (scheme-independent) leading-order (LO) unpolarized and polarized 
splitting functions read
\beq
  \delta_{\,\rm qq}^{\,(0)} = 0 \;\; , \quad
  \delta_{\,ik}^{\,(0)} =\, \mbox{ const}\, \cdot (1\!-\!x)^2 \,+\:\ldots  
  \quad \mbox{for} \quad ik = {\rm qg,\: gq,\: gg} \; .
\eeq


\begin{figure}[ht]
\vspace*{-2mm}
\centerline{\hspace*{-1mm}\epsfig{file=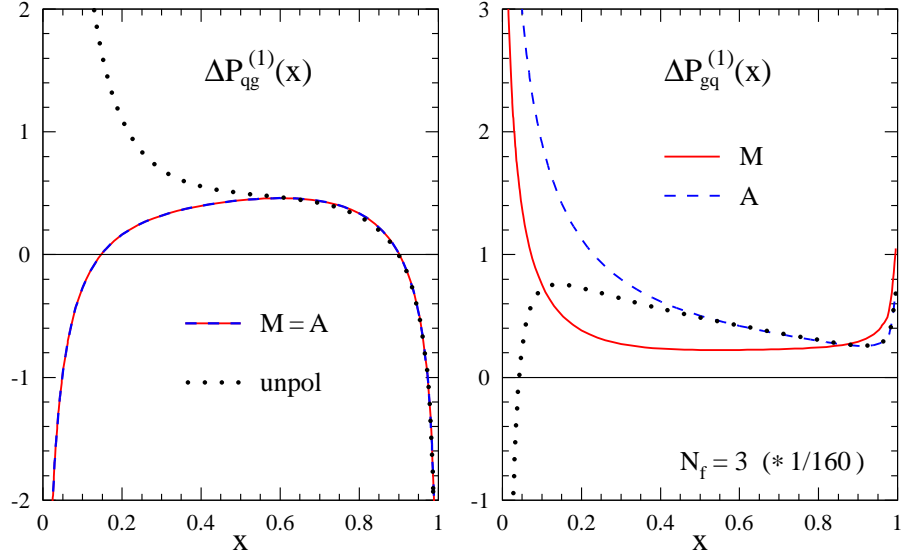,width=12.5cm}}
\vspace{-3mm}
\caption{\label{dPij-NLO} 
The two-loop (NLO) splitting functions 
$\Delta P^{\,(1)}_{i\neq k}(x)$, compared to their unpolarized counterparts.
The results are shown as published in Refs.~\cite{MvN,WVdP1} (`M') and after
an including an additional term 
$\,z_{\,\rm gq}^{(1)} = -\Delta P^{\,(0)}_{\,\rm gq}$
in the transformation from the Larin scheme (`A'), which removes all 
$(1-x)^{\,0,\,1}$ terms from $\,\delta_{\,\rm gq}^{\,(1)}$.
}
\vspace*{-2mm}
\end{figure}
 

\noindent
The corresponding NLO results are, in the standard~version (denoted by 
`M' below) of \MSbar\ \cite{MvN,WVdP1},
\bea
  \delta_{\,ik}^{\,(1)} &=&
  {\cal O} ( (1\!-\!x)^a )
  \quad \mbox{ for} \quad ik = {\rm qq,\, gg}\:
  (\mbox{with }\,a\!=\!1),\; {\rm qg}\:(\mbox{with }\,a\!=\!2)
\\
  \delta_{\,\rm gq}^{\,(1)} &=&
  8\, \* \cf \* (\ca\!-\!\cf)\, \* (2\!-\!x)\,  \* \ln (1\!-\!x)
  \,+\, 4\, \* \cf \* \beta_0^{} - 6\, \* \cfs
\nn \\[-0.5mm] & & \!\!\!
  \,+\, ( 20/3\, \* \cf \ca + 2\, \* \cfs
        - 8/3\, \* \cf \nf ) \* (1\!-\!x)
  \,+\, {\cal O} ( (1\!-\!x)^2 ) \; .
\label{dPgq1xto1}
\eea
The question arises whether these $(1-x)^{\,0}$ and $(1-x)^{\,1}$ terms are a
physical feature or a scheme artifact. 
Flavour-singlet physical evolution kernels for structure functions in DIS,
cf.~Refs.~\cite{FP82,SMVV}, 
\beq
  \frac{d F}{d \ln Q^{\,2}} \:=\:
  \frac{d\, C}{d \ln Q^{\,2}}\, f \, + \, C P f \; = \;
  \big( \beta(\ar)\, \frac{d\, C}{d \ar}\, + C P \big) C^{\,-1} F
  \; = \; K F \; ,
\eeq
if available for corresponding quantities, can provide insight on this 
question.


\section{\boldmath {\normalsize $\ass^3$} contributions via {$g_1^{}$}
(at all $N\,$), and graviton-exchange~DIS (for fixed $N$-values)}

\vspace*{-1mm}
\noindent
Following Refs.~\cite{Mom3loop1,Mom3loop2,mvvPns,mvvPsg,mvvF2L}, 
our third-order calculation of polarized DIS proceeds via the optical theorem, 
which relates probe$\,$($q$)-parton$\,$($p$) total cross sections 
(with $Q^{\,2} = - q^{\,2} > 0$ and $p^{\,2}$ = 0) \linebreak
to forward amplitudes, and a dispersion relation in $x\,$, which 
provides the $N$-th Mellin moment 
\beq
  A^N \:=\: {\textstyle \int}_{\!0}^1 \, dx \; x^{\,N-1} A(x)
\eeq
from the coefficient of $(2p\!\cdot\! q)^N$. The unpolarized case was first
computed at even $N \leq 10$ in the mid 1990s in 
Refs.~\cite{Mom3loop1,Mom3loop2}, using the {\sc Mincer} program for 
three-loop self-energy integrals \cite{MINCER}. The corresponding all-$N$
and all-$x$ expressions were derived by us ten years ago 
\cite{mvvPns,mvvPsg,mvvF2L}.

\vspace*{0.5mm}
A brief account of the extension of the latter calculations to the polarized 
structure function \linebreak $g_{1\!}^{}$ was presented at Loops \& Legs 2008 
\cite{mvvLL2008}, where we focused on the resulting expressions for 
$\Delta P_{\rm qq}^{\,(2)}$ and $P_{\,\rm qg}^{(2)}$ which can by extracted
from the $\ep^{\,-1}$ poles of the unfactorized structure functions.  


\noindent
The resulting $\cfs\nf$ contribution to the latter function, in the standard 
$M$ scheme, is given by
\bea
  && \hspace*{-6mm} 
  \frct{1}{8}\,\Delta P^{\,(2)}_{\,\rm qg}(N) \big|_{\cfs\nf}
  \: = \;
          2\, \* \colour4colour{ \Delta p_{\rm qg} } \* (
          - \, \S(-4)
          + 2\, \* \Ss(-2,-2)
          + 4\, \* \Ss(1,-3)
          + 2\, \* \Ssss(1,1,1,1)
          - \Sss(1,1,2)
          - 5\, \* \Sss(1,2,1)
\quad \nn \\[-1.8mm] && 
 \hspace*{3.65cm}
\label{dPqg2cf2nf}
          + \, 4\, \* \Ss(1,3)
          + 2\, \* \Ss(2,-2)
          - 6\, \* \Sss(2,1,1)
          + 6\, \* \Ss(2,2)
          + 7\, \* \Ss(3,1)
          - 3\, \* \S(4)
          )
\nn \\[0.5mm] && \hspace*{-3mm}
       - \, 3 \* \colour4colour{ \z3 } \, \*  (
            2\, \* \DNn2
          + 4\, \* \DNp2
          - 9\, \* \DNnO
          + 12\, \* \DNpO
          )
       + 4\, \* \colour4colour{ \S(-3) }\, \*  (
            \DNn2
          - 2\, \* \DNnO
          + 2\, \* \DNpO
          )
       + 8\, \* \colour4colour{ \Ss(1,-2) }\, \*  (
            2\, \* \DNp2
          - \DNnO
          + \DNpO
          )
\nn \\[-0.5mm] &&
       - 2\, \* \colour4colour{ \Ss(2,1) }\, \*  (
            4\, \* \DNn2
          + 2\, \* \DNp2
          - 11\, \* \DNnO
          + 11\, \* \DNpO
          )
       + \colour4colour{ \Sss(1,1,1) }\, \*  (
            5\, \* \DNn2
          - 2\, \* \DNp2
          - 21/2\, \* \DNnO
          + 12\, \* \DNpO 
          )
\nn  \\[-0.5mm] &&
       - 2\, \* \colour4colour{ \Ss(1,2) }\, \*  (
            2\, \* \DNn2
          - 2\, \* \DNp2
          - 5\, \* \DNnO
          + 5\, \* \DNpO
          )
       + 2\, \colour4colour{ \* \S(3) }\, \*  (
            3\, \* \DNn2
          + 6\, \* \DNp2
          - 11\, \* \DNnO
          + 11\, \* \DNpO
          )
 \\[0.8mm] && \hspace*{-3mm}
       + \, 2\, \* \colour4colour{ \S(-2) }\, \*  (
            8\, \* \DNp3
          - 5\, \* \DNn2
          - 6\, \* \DNp2
          + 10\, \* \DNnO
          - 9\, \* \DNpO
          )
       - \colour4colour{ \Ss(1,1) }\, \*  (
            10\, \* \DNn3
          + 6\, \* \DNp3
          - 35/2\, \* \DNn2
          - 5\, \* \DNp2
\nn \\[-0.5mm] &&
          + 29\, \* \DNnO
          - 36\, \* \DNpO
          )
       + 2\, \* \colour4colour{ \S(2) }\,  \*  (
            4\, \* \DNn3
          + 6\, \* \DNp3
          - 10\, \* \DNn2
          - 4\, \* \DNp2
          + 17\, \* \DNnO
          - 22\, \* \DNpO
          )
       \,\,
          - 6\, \* \colour4colour{ \DNppO }\, \*  ( \S(-2) + 1 ) 
\nn \\[0.8mm] && \hspace*{-3mm}
       + \, \colour4colour{ \S(1) }\, \*  (
            7\, \* \DNn4
          + 4\, \* \DNp4
          - 43/2\, \* \DNn3
          - 15\, \* \DNp3
          + 99/2\, \* \DNn2
          + 18\, \* \DNp2
          - 78\, \* \DNnO
          + 329/4\, \* \DNpO
          )
          + 32\, \* \DNp5
\nn \\[-0.5mm] && \hspace*{-3mm}
          - 15/2\, \* \DNn4
          - 3\, \* \DNp4
          + 59/8\, \* \DNn3
          + 53/4\, \* \DNp3
          + 77/8\, \* \DNn2
          + 213/8\, \* \DNp2
          - 1357/32\, \* \DNnO
          + 777/16\, \* \DNpO
\nn 
\eea
in terms of $D_k = (N\!+k)^{-1}$ and $\Delta p_{\rm qg}= 2\,D_1^{} - D_0^{}$,
with all harmonic sums \cite{HSums} at argument $N$. 

\vspace*{0.3mm}
This result shows some interesting features. The weight-4 sums in the first
two rows have the same coefficient in the unpolarized case of 
Ref.~\cite{mvvPsg}, where $\Delta p_{\rm qg}$ is replaced by its 
counterpart $p_{\rm qg}$. The lower-weight denominator structure is simpler
in the present case, with only two terms with $D_2^{}$ (third line from below) 
which do not lead to additional denominator primes at odd values of $N$. As in
previous results in massless QCD, Eq.~(\ref{dPqg2cf2nf}) does not include sums
with index $-1$. The large-$N$ suppression of $\,\delta_{\,\rm qg}^{\,(2)}$
by two powers of $1/N$ holds separately for each harmonic sum. Finally the
coefficients $D_{0,1}^{\,5}$, $D_1^{\,4\,}$ and $\Sss(1,1,1)$ are 
predictable in terms of $x\! \ra\! 0\,$ and $x\! \ra\! 1$ knowledge, i.e., by
Ref.~\cite{BVpol} and by extending Ref.~\cite{AV11xto0}, see also Ref.~\cite
{avLL2012}, and Ref.~\cite{SMVV} to the present case.


\vspace*{2mm}
The lower-row splitting functions $\,\Delta P_{\,\rm gq}^{\,(2)}$ and
$\,\Delta P_{\,\rm gg}^{\,(2)}$ enter standard (electroweak gauge-boson 
exchange) DIS only at order $\ass^4$. 
Hence an additional probe directly coupling to gluons is required. Following
Ref.~\cite{FP82}, the computation of $F_2^{}$ has been complemented by DIS via
a scalar $\phi$ with a $\phi\:\! G^{\,\mu\nu\!} G_{\mu\nu}$ coupling to gluons,
i.e., the Higgs boson in the heavy-top limit, in 
Refs.~\cite{Mom3loop2,mvvPsg}.

\vspace{0.3mm}
In the polarized case a non$\,$-$\,$(pseudo)$\,$scalar probe is
required, in contrast to our statement in the penultimate paragraph of
Ref.~\cite{mvvLL2008}, which was based on an incorrectly simplified diagram
database. One way to address this issue would be to extent the calculations
to a supersymmetric case, as done in the context of NNLO antenna functions
in Ref.~\cite{GdGGN05}. Instead we consider graviton-exchange DIS, as 
described in Ref.~\cite{LamLi}, see also Ref.~\cite{SVgDIS}, which provides
five relevant structure functions, $H_k$, $k = 1\!-\!4, 6$, that can be
combined to provide unpolarized and polarized analogues of the system
$(F_2, F_\phi)$, plus an analogue of the standard longitudinal structure
function $F_{L\,}^{}$.

\vspace{0.3mm}
A major drawback of this approach is that it leads to a very large number of
higher tensor integrals, far beyond those 
tabulated during the 
calculation of $F_{2}^{}$ and $F_\phi$ \cite{mvvPns,mvvPsg,mvvF2L} and its
later extension to $g_1^{}$ \cite{mvvLL2008}. We have therefore decided to 
(first) fall back to fixed-$N$ calculation using {\sc Mincer} \cite{MINCER}, 
for which we have improved our diagram management and, in particular, the 
high-$N$ efficiency of the {\sc Mincer} program, see Ref.~\cite{jvLL2014}. 
These improvements have allowed us to calculate polarized graviton-exchange 
DIS at the third order completely for the 12 odd moments $3 \leq N \leq 25$.
The first moments are directly accessible neither in our calculation nor
via operator matrix elements~\cite{LamLi}. 

\vspace{0.5mm}
The calculations were performed on computers at
DESY-Zeuthen (mainly for {\sc Mincer} development), NIKHEF (hardest diagrams
at highest values of $N$) and the {\tt ulgqcd} cluster in Liverpool (bulk
production, using more than 200 cores), using the latest version of TFORM
\cite{FORM}.


\noindent
As an example, we here show the calculated moments of the $\cft$ part of
$\,\Delta P_{\rm gq}^{\,(2)}$ in the Larin scheme.
\\[2mm] \noindent
{\footnotesize
~N = ~3: ~$ 186505/7776 $
\\[0.2mm]
~N = ~5: ~$ 9473569/3037500 $
\\[0.2mm]
~N = ~7: $ -509428539731/193616640000 $
\\[0.2mm]
~N = ~9: $ -266884720969207/56710659600000 $
\\[0.2mm]
~N = 11: $ -3349566589170829651/608887229282640000 $
\\[0.2mm]
~N = 13: $ -751774767290148022507/130490947198868256000 $
 \hspace{6cm} {\normalsize (2.3)}
\\[0.2mm]
~N = 15: $ -23366819019913026454180147/4047226916198744678400000 $
\\[0.2mm]
~N = 17: $ -305214227818628090680174170947/53873282508311259589115520000 $
\\[0.2mm]
~\mbox{N = 19:~$ -570679648684656807578199791973487/103793635967590259537308862400000 $}
\\[0.2mm]
~\mbox{N = 21:~$ -2044304092089235762279148843319979/385456787045956248050132280576000 $}
\\[0.2mm]
~N = 23:~$ -289119840113761409530260333250139823739/
56707019270988141152999601215071395840 $
\\[0.2mm]
~N = 25:~$ -1890473255283802937678830745102921869938637/
386426908528565021863360305851160000000000 $
}


\vspace{1.5mm}
Returning to the large-$x$ limit, we note that the unpolarized structure 
functions $H_{\,\bar{2}}^{}$ (LO: quarks, due to forming a suitable linear
combination of $H_{\,2}^{}$ and $H_{\,3}^{}$) and $H_{\,3}^{}$ (LO: gluons, 
from the outset) and their polarized counterparts $H_{\,\bar{4}}^{}$, 
$H_{\,6}^{}$ form a set of quantities as mentioned at the end of Section~1. 
\linebreak
Comparing the NLO evolution kernel $K_{\,3\bar{2}}^{\,(1)}$ and 
$K_{\,6\bar{4}}^{\,(1)}$, which we have calculated at all $N\,/\,$all $x$, we 
can conclude that the large-$x$ behaviour of $\,\Delta P_{\rm gq}^{\,(1)}$ 
of Refs.~\cite{MvN,WVdP1} discussed above is not physical.

\vspace{0.5mm}
Consequently one may expect the existence of a simple additional NNLO 
transformation that restores also the $1/N^{\,2}$ suppression of 
$\delta_{\,\rm gq}^{\,(2)}(N) =  \,P_{\rm gq}^{\,(2)}(N) - \Delta
P_{\rm gq}^{\,(2)}(N)$. As shown in Fig.~2, where all non-$\nf$ and 
$\nf^{\!\!\!1}$ colour factors have been combined for brevity, this 
expectation appears to be justified. Hence the three-loop analogue of 
Eq.~(\ref{dPgq1xto1}) can be predicted from lower-order information. 


\begin{figure}[ht]
\vspace*{-2mm}
\centerline{\hspace*{-1mm}\epsfig{file=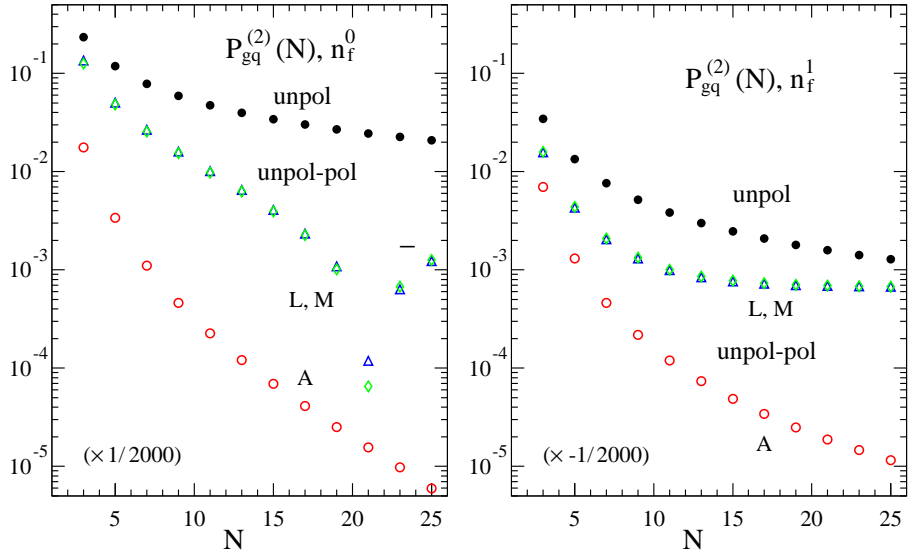,width=12.5cm}}
\vspace{-2mm}
\caption{\label{dPgq-NNLO}
The moments of the three-loop (NNLO) splitting functions 
$\Delta P^{\,(2)}_{\,\rm gq}$ in QCD determined using the {\sc Mincer}
program for gravition-exchange DIS. The results are shown separately for the 
$\nf^{\!\!\!0}$ and $\nf^{\!\!\!1}$ part in the Larin scheme (`L'), the 
standard \MSbar\ scheme according to Ref.~\cite{MSvN} (`M') and with a NNLO 
additional term $\,z_{\,\rm gq}^{(2)} = -\frac{1}{2} \Delta P^{\,(1)L}_
{\,\rm gq}$ in the transformation from the Larin scheme to \MSbar\ (`A').
}
\vspace*{-4mm}
\end{figure}


\section{All-$N$ expressions, using end-point knowledge and number-theory 
tools}

\vspace*{-1mm}
\noindent
We illustrate the determination of the all-$N$ expressions for the critical 
$\nf^{\!\!\!0}$ parts of $\Delta P_{\rm gq}^{\,(2)}(N)$. 
Analogous to Eq.~(\ref{dPqg2cf2nf}), the coefficients of the weight-4 sums are 
fixed by the unpolarized case. This leaves 2$\times$32 coefficients of sums at 
weight three and below combined with powers of $N^{\,-1}$ and 
\mbox{$(N\!+\!1)^{\,-1}$}, plus up to 11 sums combined with $(N\!-\!1)^{\,-1}$.
Of these 75 unknowns, the 24 coefficients of $D_0^{\,1}$ and $D_1^{\,1}$ 
can be eliminated using the empirical $1/N^{\,2}$ large-$N$ suppression of 
$\delta_{\,\rm gq}^{\,(2)}(N)$ in the $A$-scheme, and a further 6 from
small-$x$ and large-$x$ constraints as discussed below Eq.~(\ref{dPqg2cf2nf}).

\vspace*{0.5mm}
We have developed FORM tools which analyze the prime decomposition of the 
calculated moments and facilitate the derivation of relations between the 
remaining coefficients (which are all integer if suitably normalized) using 
the Chinese remainder theorem. These have
proved sufficient, sometimes together with a brute-force scan of a few 
remaining variables, to solve simpler cases. It~is however rather hard to get 
more than about ten relations for the difficult $\nf^{\!\!\!0}$ parts of 
$\Delta P_{\rm gq}^{\,(2)}(N)$.

\vspace*{0.5mm}
Motivated by Ref.~\cite{Veliz}, we have turned to professional number-theory 
tools for these cases, in particular the program provided at~
 {\tt www.numbertheory.org/php/axb.html}~
which 
{\it `Solves a system of linear Diophantine equations using the Hermite normal 
form of an integer matrix via the Havas-Majewski- Matthews LLL-based algorithm.
\dots$\:\!$ . We find \dots the solutions X with minimal length, using a 
modification of the Fincke-Pohst algorithm'} \cite{axbCode}.

\vspace*{0.5mm}
Since that algorithm looks for short vectors, it is best for our purposes to 
eliminate, say, six `unpleasant' coefficients, in particular those of the 
low-weight combinations 
$D_{0}^{\,2}$, $D_{1}^{\,2}$, $D_{0}^{\,2} S_1$, $D_{1}^{\,2} S_1$, 
using the moments $3 \leq N \leq 13$, and to use the above program for the 
remaining six equations. 


\vspace*{1.5mm}
Using the moments shown on the previous page, this procedure leads to
the $M$-scheme result
\bea
  && \hspace*{-6mm} 
  \frct{1}{8}\,\Delta P^{\,(2)}_{\,\rm gq}(N) \big|_{\cft}
  \: = \;
          2\, \* \colour4colour{ \Delta p_{\rm gq} } \* (
          - \, \S(-4)
          + 6\, \* \Ss(-2,-2)
          + 4\, \* \Ss(1,-3)
          + 2\, \* \Ssss(1,1,1,1)
          + \Sss(1,1,2)
\nn \\[-1mm] &&
 \hspace*{3.45cm}
          + \,3\, \* \Sss(1,2,1)
          - 3\, \* \Ss(1,3)
          + 2\, \* \Ss(2,-2)
          + 2\, \* \Sss(2,1,1)
          - 2\, \* \Ss(2,2)
          )
\nn \\[0.5mm] && \hspace*{-3mm}
       + \, 6\, \* \colour4colour{ \z3\, \* \Delta p_{\rm gq} }
         ( 2\, \* \S(1) - 3 )
       - \, 4 \, \* \colour4colour{ \S(-3) }\, \*  (
            2\, \* \DNn2
          - \DNnO
          + \DNpO
          )
       - 8 \, \* \colour4colour{ \Ss(1,-2) }\, \*  (
            \DNp2
          - 2\, \* \DNnO
          + 2\, \* \DNpO
          )
\nn \\[-0.5mm] &&
       + \, \colour4colour{ \Sss(1,1,1) }\, \*  (
            2\, \* \DNn2
          - 5\, \* \DNp2
          - 6\, \* \DNnO
          - 3/2\, \* \DNpO
          )
       - 2\, \* \colour4colour{ \Ss(1,2) }\, \*  (
            \DNp2
          + 4\, \* \DNnO
          - \DNpO
          )
\nn \\[-0.5mm] &&
       - \, \colour4colour{ \Ss(2,1) }\, \*  (
            4\, \* \DNn2
          + 4\, \* \DNp2
          - 4\, \* \DNnO
          + 7\, \* \DNpO
          )
       + \colour4colour{ \S(3) }\,  \*  (
            2\, \* \DNn2
          + \DNp2
          + 6\, \* \DNnO
          - 3/2\, \* \DNpO
          )
\\[0.8mm] && \hspace*{-3mm}
       \, - \, \colour4colour{ \S(-2) }\, \*  (
            8\, \* \DNp3
          + 4\, \* \DNn2
          + 18\, \* \DNp2
          - 26\, \* \DNnO
          + 24\, \* \DNpO
          )
       + 2 \, \* \colour4colour{ \S(2) }\, \*  (
            \DNp3
          + 2\, \* \DNp2
          + 10\, \* \DNnO
          - 4\, \* \DNpO
          )
\nn \\[-0.5mm] &&
       \, - \, \colour4colour{ \Ss(1,1) }\, \*  (
            6\, \* \DNn3
          + 6\, \* \DNp3
          + 4\, \* \DNn2
          + 5\, \* \DNp2
          + 2\, \* \DNnO
          - 7/4\, \* \DNpO
          )
        \,\, - \,\, 6 \, \* \colour4colour{ \DNmO }\, \*  ( \S(-2) + 1 )
\nn \\[0.8mm] && \hspace*{-3mm}
       \, - \, \colour4colour{ \S(1) }\, \*  (
            6\, \* \DNn4
          + 7\, \* \DNp4
          + 4\, \* \DNn3
          + 23/2\, \* \DNp3
          - 27/2\, \* \DNn2
          + 39/4\, \* \DNp2
          - 8\, \* \DNnO
          + 23/4\, \* \DNpO
          )
\nn \\[0.8mm] && \hspace*{-3mm}
          \, - \, 8\, \* \DNn5
          - 12\, \* \DNp5
          + 23\, \* \DNn4
          - 28\, \* \DNp4
          - 39/4\, \* \DNn3
          - 427/8\, \* \DNp3
          - 341/8\, \* \DNn2
          - 767/8\, \* \DNp2
\nn \\[-0.5mm] &&
          + 2427/16\, \* \DNnO
          - 4547/32\, \* \DNpO
\nn 
\eea
with $\Delta p_{\rm gq} = 2\,D_0 - D_1$ and, again, 
$D_k = (N\!+k)^{-1}$ and all harmonic sums taken at argument $N$.
The corresponding expressions for the $\cf \cas$ and $\cfs \ca$ parts are
somewhat lengthier; while the $\nf$-dependent terms are much simpler and do 
not require the $N\!=\!25$ moment. The determination of the all-$N$ result
for the NNLO gluon-gluon splitting function $\Delta P_{\,\rm gq}^{\,(2)}$
proceeded in an analogous manner; finding the all-$N$ form of its $\cat$ part 
was the overall most difficult task. 


\vspace*{1.5mm}
While it is easy to recognize, by looking at the pattern of the coefficients,
whether or not the correct all-$N$ form is returned by the solution of a
Diophantine system, it is necessary to validate the results. 
For this purpose the results are Mellin-inverted to $x$-space expressions 
$\Delta P_{\,\rm gq}^{\,(2)}(x)$  and $\Delta P_{\,\rm gg}^{\,(2)}(x)$ in
terms of harmonic polylogarithms \cite{HPols}, from which arbitrary moments can
be determined. The results can thus be compared to additional moments
calculated using {\sc Mincer}, such~as 
%
{\small
\bea
  - \Delta P^{\,(2)}_{\,\rm gq}(N\!=\!27) &=& 
  4609770383587605432813291530849726335264810727/ \nn \\[-1mm]
& & 
  982934508627216318966565777854990940800000000\: \* \cft 
 \, + \:\ldots
\eea
} 
with \quad {\small
Total execution time: 256 874 306.6 sec.
Maximum disk space: 1 261 024 031 636 bytes.
}
\\[2mm]
Further high-$N$ checks have been performed for 
$\,\Delta P^{\,(2)}_{\,\rm gq}(N\!=\!29)\,$ in the planar limit
$\,\ca\! -\! 2\cf \ra 0\,$ at $\nf = 0$, which combines the three difficult 
all-$N$ expressions, and for the crucial $\cat$ parts of 
$\,\Delta P^{\,(2)}_{\,\rm gg}(N)$ at $N=27$ and $N=29$. The functions 
$\Delta P_{\,\rm gq}^{\,(2)}(x)$ and $\Delta P_{\,\rm gg}^{\,(2)}(x)$ pass all 
these tests.

\vspace*{1.5mm}
Finally these $x$-space expressions also facilitates the determination of 
the first moments,
\bea
\label{dPgq2N1}
   \Delta P^{\,(2)}_{\,\rm gq}(N\!=\!1) \! &\:\:=\:\:&
              \frct{1607}{12} \, \* \cf \* \, \cas
         \,-\, \frct{461}{4} \, \* \cfs \* \, \ca
         \,+\,  \frct{63}{2} \* \cft
         \,+\, \Big( \,\frct{41}{3} - 72\*\z3 \Big) \,\* \cf \, \* \ca \, \* \nf
\nn \\ & & 
         \,-\, \Big( \,\frct{107}{2} - 72\* \z3 \Big) \, \* \cfs \* \, \nf
         \,-\, \frac{13}{3} \, \* \cf \* \nfs 
\;\; , \\[2.5mm] \Delta P^{\,(2)}_{\,\rm gg}(N\!=\!1) \! &\:\:=\:\:&
              \frct{2857}{54} \, \* \cat
         \,-\, \frct{1415}{54} \, \* \cas \* \, \nf
         \,-\, \frct{205}{18} \, \* \cf \, \* \ca \, \* \nf
         \,+\, \cfs \, \* \nf 
         \,+\, \frct{79}{54} \, \ca \, \* \nfs 
         \,+\, \frct{11}{9} \, \cf \, \* \nfs
\nn \\[1mm] &\:\:=\:\: & 
  \beta_{\,2}^{\,\mbox{\scriptsize \MSbar}} 
\label{dPgg2N1}
\; . 
\eea
The agreement, for all six colour factors, of 
$\Delta P^{\,(2)}_{\,\rm gg}(N\!=\!1)$ with the NNLO contribution 
\cite{beta2} to the $\beta$-function of QCD in the \MSbar\ scheme provides 
another strong check of our results. 


\vspace*{1.5mm}
The new splitting functions $\Delta P^{\,(2)}_{\,\rm gq}(x)$ and
$\Delta P^{\,(2)}_{\,\rm gg}(x)$ are shown in Fig.~3. 
As in the previous figures, the curves are scaled such that the results are 
approximately converted from the small parameter $\ar = \as/(4 \pi)$ in 
Eq.~(\ref{Pexp}) to an expansion in $\as$. In Fig.~4 the impact of these
results on the evolution is illustrated for a sufficiently realistic
model input \cite{NumEvol} at a rather large value of $\as$.

\begin{figure}[ht]
\centerline{\hspace*{-1mm}\epsfig{file=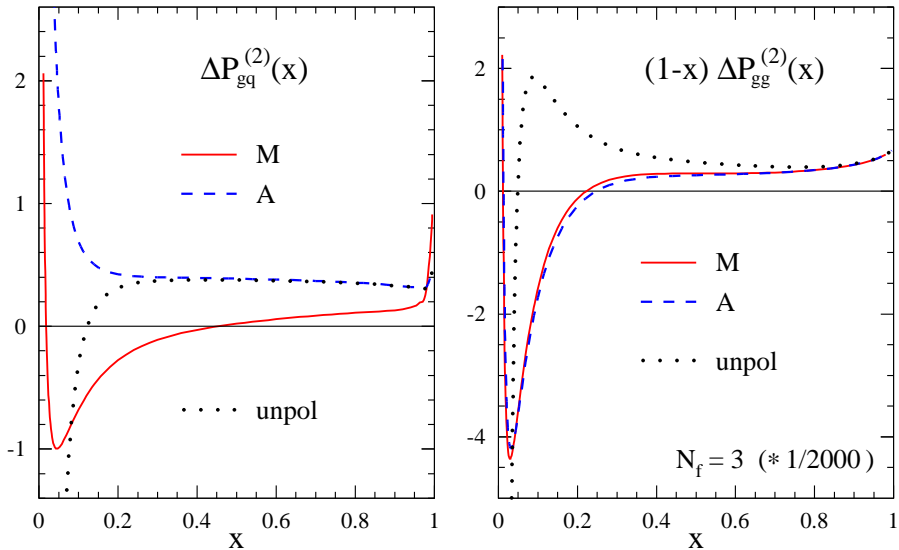,width=12.5cm}}
\vspace{-3mm}
\caption{\label{dPgi-NNLO}
The NNLO splitting functions $\Delta P^{\,(2)}_{\,\rm gq}(x)$
(left) and $\Delta P^{\,(2)}_{\,\rm gg}(x)$ (right) compared to the
corresponding unpolarized quantities.  The results are shown in the $M$ and 
$A$ schemes for three light flavours~$\nf$.
}
\vspace*{-5mm}
\end{figure}


\begin{figure}[ht]
\vspace*{-1mm}
\centerline{\hspace*{-1mm}\epsfig{file=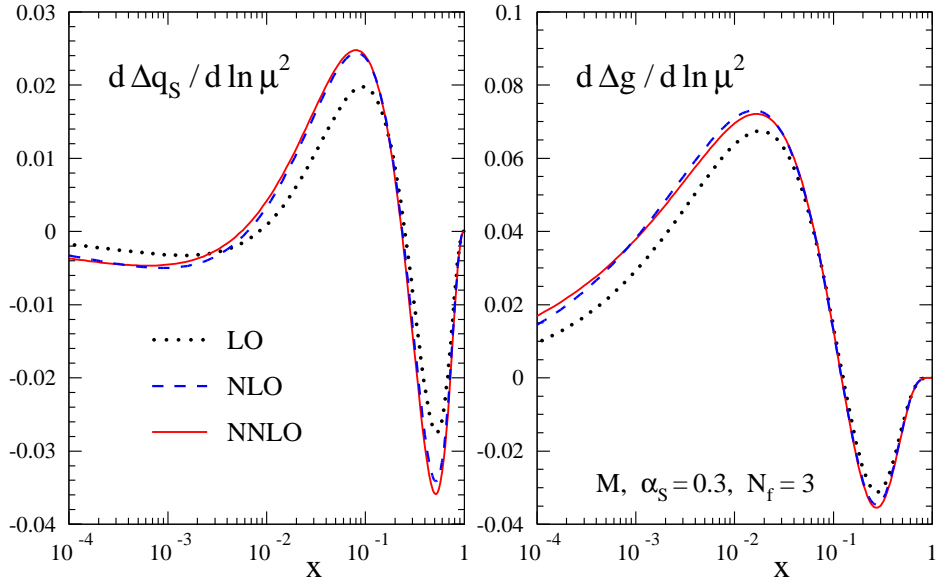,width=12.5cm}}
\vspace{-2mm}
\caption{\label{dsgq-evol}
The LO, NLO and NNLO approximations to the scale derivatives of the polarized
singlet quark (left) and gluon (right) distributions in the standard 
version (`M') of the \MSbar\ scheme \cite{MvN,WVdP1,MSvN}, for the (order-independent) benchmark initial
distributions of Refs.~\cite{NumEvol} at a low scale $\mu^2$ with 
$\as(\mu^2) = 0.3$.
}
\vspace*{-2mm}
\end{figure}

\section{Summary and outlook: more checks and calculations}

\vspace*{-1mm}
\noindent
We have finally, 10 years after publishing their unpolarized counterparts
\cite{mvvPns,mvvPsg}, derived all NNLO helicity-difference splitting functions
$\Delta P^{\,(2)}_{ij}(x)$. The last part, the lower row
$\Delta P^{\,(2)}_{\,\rm gq}$ and $\Delta P^{\,(2)}_{\,\rm gg}$ of the
flavour-singlet matrix, has been obtained by a combination of brute-force
computations using {\sc Mincer} \cite{MINCER}, insights into the structure of
these functions, and number-theory tools \cite{axbCode}.

\vspace*{0.5mm}
The three-loop {\sc Mincer} computations of graviton-exchange DIS 
\cite{LamLi} have also been performed for the unpolarized case and, also to 
very high values of the Mellin moment $N$, for the upper row for which we had 
calculated the all-$N$ results before \cite{mvvLL2008}. 
The resulting agreement with the corresponding splitting functions provides 
checks of our treatment of graviton-exchange DIS and of the {\sc Mincer} code 
as modified for much better large-$N$ performance.

\vspace*{0.5mm}
Our results agree with all previous partial results 
-- if interpreted properly; in particular, the leading small-$x$ terms of 
Ref.~\cite{BVpol} apply to the NNLO physical kernels in the off-diagonal cases,
not to the corresponding \MSbar\ splitting functions -- 
and expectations for the high-energy and threshold limits, the first moments 
of $\Delta P_{\,\rm gg}$ and the leading large-$\nf$ contributions~%
\cite{LargeNf}.

\vspace*{0.5mm}
As for the unpolarized case, the numerical effects of these NNLO contributions
are small down to low values of $x$ after the convolution with realistic
quark and gluon initial distributions. \linebreak 
The published version of the \MSbar\ 
scheme, defined by the transformation correcting for the use of, e.g., the
Larin scheme for $\gamma_5^{}$ in dimensional regularization, is somewhat
unphysical for $x\! \ra\! 1$ already at NLO. However this does not appear to 
be a practically relevant problem, hence we see no reason to advocate a change 
of the scheme after almost 20 years of NLO data analyses.

\vspace*{0.5mm}
Nevertheless, a re-calculation of the critical NNLO transformation quantity 
$z_{\rm ps}^{\,(2)}$ (and a check of $z_{\,\rm gq}^{\,(n)}\,=\,0$) would be
worthwhile. In fact, its extension to the third order would suffice to fix 
the N$^{3}$LO quark coefficient function for $g_1^{}$, as we obtained the 
Larin-scheme result some years ago.


\section*{Acknowledgements}

\vspace*{-1mm}
\noindent
We would like to thank John Gracey for useful discussions.
This work has been supported by 
the UK {\it Science \& Technology Facilities Council}$\,$ (STFC) under grant 
number ST/G00062X/1,
the \mbox{German}
{\it Bundesministerium f\"ur Bildung und Forschung} through contract 05H12GU8,
the {\it European Research Council}$\,$ (ERC) Advanced Grant no.~320651, 
{\it HEPGAME}, 
and by the European Commission through contract PITN-GA-2010-264564 
({\it LHCPhenoNet}$\,$). 
We are particularly grateful for the opportunity to use a substantial part of 
the {\tt ulgqcd} computer cluster in Liverpool which was funded by STFC under 
grant number ST/H008837/1.

\end{document}